\documentclass[twocolumn,showpacs,superscriptaddress,preprintnumbers]{revtex4-1}
\usepackage{amsfonts, amsmath, amssymb, bm, bbm, graphicx, epsfig, epstopdf}
\usepackage{dcolumn}
\usepackage{textcomp}
\usepackage{hyperref}
\usepackage{color}
\usepackage{natbib}
\usepackage[utf8]{inputenc}    
\usepackage[normalem]{ulem}
\usepackage{mathtools}
\usepackage{epstopdf}

\def\opone{\leavevmode\hbox{\small1\kern-3.8pt\normalsize1}}

\newcommand{\ket}[1]{\ensuremath{\left|{#1}\right\rangle}}
\newcommand{\bra}[1]{\ensuremath{\left\langle{#1}\right |}}

\begin{document}

\title{Quantum teleportation across a metropolitan fibre network}

\author{Raju Valivarthi}
\affiliation{Institute for Quantum Science and Technology, and Department of Physics \& Astronomy, University of Calgary, 2500 University Drive NW, Calgary, Alberta T2N 1N4, Canada}

\author{Marcel.li {Grimau Puigibert}}
\affiliation{Institute for Quantum Science and Technology, and Department of Physics \& Astronomy, University of Calgary, 2500 University Drive NW, Calgary, Alberta T2N 1N4, Canada}

\author{Qiang Zhou}
\affiliation{Institute for Quantum Science and Technology, and Department of Physics \& Astronomy, University of Calgary, 2500 University Drive NW, Calgary, Alberta T2N 1N4, Canada}

\author{Gabriel H. Aguilar}
\affiliation{Institute for Quantum Science and Technology, and Department of Physics \& Astronomy, University of Calgary, 2500 University Drive NW, Calgary, Alberta T2N 1N4, Canada}

\author{Varun B. Verma}
\affiliation{National Institute of Standards and Technology, Boulder, Colorado 80305, USA}

\author{Francesco Marsili}
\affiliation{Jet Propulsion Laboratory, California Institute of Technology, 4800 Oak Grove Drive, Pasadena, California 91109, USA}

\author{Matthew D. Shaw}
\affiliation{Jet Propulsion Laboratory, California Institute of Technology, 4800 Oak Grove Drive, Pasadena, California 91109, USA}

\author{Sae Woo Nam}
\affiliation{National Institute of Standards and Technology, Boulder, Colorado 80305, USA}

\author{Daniel Oblak}
\affiliation{Institute for Quantum Science and Technology, and Department of Physics \& Astronomy, University of Calgary, 2500 University Drive NW, Calgary, Alberta T2N 1N4, Canada}

\author{Wolfgang Tittel*}
\affiliation{Institute for Quantum Science and Technology, and Department of Physics \& Astronomy, University of Calgary, 2500 University Drive NW, Calgary, Alberta T2N 1N4, Canada}

\date{\today }

\maketitle




\textbf{If a photon interacts with a member of an entangled photon pair via a so-called Bell-state measurement (BSM), its state is teleported over principally arbitrary distances onto the second member of the pair \cite{Bennett93}. Starting in 1997 \cite{Boschi98, Zeilinger_telep, furusawa}, this puzzling prediction of quantum mechanics has been demonstrated many times \cite{Pirandola15}; however, with one very recent exception \cite{Hensen15}, only the photon that received the teleported state, if any, travelled far while the photons partaking in the BSM were always measured closely to where they were created. Here, using the Calgary fibre network, we report quantum teleportation from a telecommunication-wavelength photon, interacting with another telecommunication photon after both have travelled over several kilometres in bee-line, onto a photon at 795~nm wavelength. This improves the distance over which teleportation takes place from 818~m to 6.2~km. Our demonstration establishes an important requirement for quantum repeater-based communications \cite{Sangouard_review} and constitutes a milestone on the path to a global quantum Internet \cite{kimble}.}


While the possibility to teleport quantum states, including the teleportation of entangled states, has been verified many times using different physical systems (see Ref. [\citenum{Pirandola15}] for a recent review), the maximum distance over which teleportation is possible --- which we define to be the spatial separation between the BSM and the photon, at the time of this measurement, that receives the teleported state  --- has so far received virtually no experimental attention. To date, only two experiments have been conducted in a setting that resulted in a teleportation distance that exceeds the laboratory scale \cite{Hensen15,landry}, even if in a few demonstrations the bee-line distance travelled by the photon that receives the teleported state has been much longer \cite{Ma12,Yin12}. 

The reason to stress the importance of distances is linked to the ability of exploiting teleportation in various quantum information applications. One important example is the task of extending quantum communication distances using quantum repeaters \cite{Sangouard_review}, most of which rely on the creation of light-matter entanglement, e.g. by creating an entangled two-photon state out of which one photon is absorbed by a quantum memory for light \cite{Lvovsky}, and entanglement swapping \cite{Zuckowski}. The latter shares the Bell state measurement (BSM) with standard teleportation; however, the photon carrying the state to be teleported is itself a member of an entangled pair. Entanglement swapping is therefore sometimes referred-to as teleportation of entanglement. To be useful in such a repeater, two entangled photon pairs must be created far apart, and the BSM, which heralds the existence of the two partaking photons and hence of the remaining members of the two pairs, should, for optimal performance, take place approximately halfway in-between these two locations. 

\begin{figure*}
\centering
\includegraphics[width=2 \columnwidth]{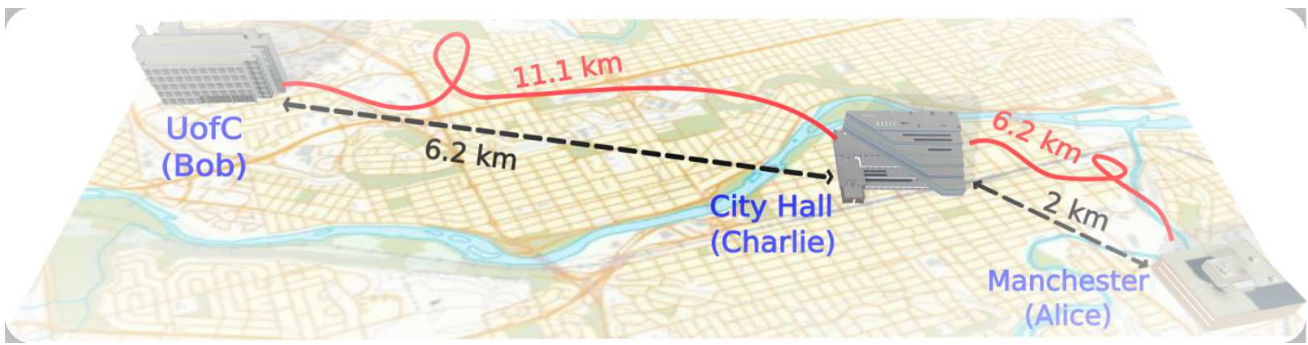}
\caption{\textbf{Aerial view of Calgary.} Alice is located in Manchester, Bob at the University of Calgary (UofC), and Charlie in a building next to City Hall in Calgary downtown. The teleportation distance --- in our case the distance between Charlie and Bob --- is 6.2 km. All fibres belong to the Calgary telecommunication network but, during the experiment, they only carry signals created by Alice, Bob or Charlie and were otherwise ``dark".
}
\label{fig:map}
\end{figure*}

Yet, due to the difficulty to guarantee indistinguishability of the two interacting photons after their transmission through long and noisy quantum channels \cite{Rubenok13}, entanglement swapping or standard teleportation in the important midpoint configuration has only been reported very recently outside the laboratory \cite{Hensen15}. This work exploited the heralding nature of the BSM for the first loophole-free violation of a Bell inequality --- a landmark result that exemplifies the importance of this configuration. However, the two photons featured a wavelength of about 637~nm, which, due to high loss during transmission through optical fibre, makes it impossible to extend the transmission distance to tens, let alone hundreds, of kilometers. In all other demonstrations, either the travel distances of the two photons were small, or they were artificially increased using fibre on spool \cite{landry, Riedmatten04,felix14}, effectively increasing travel time and transmission loss --- and hence decreasing communication rates --- rather than real separation. Here we report the first demonstration of quantum teleportation over several kilometers in the mid-point configuration and with photons at telecommunication wavelength.



\begin{figure*}
\centering
\includegraphics[width=\textwidth]{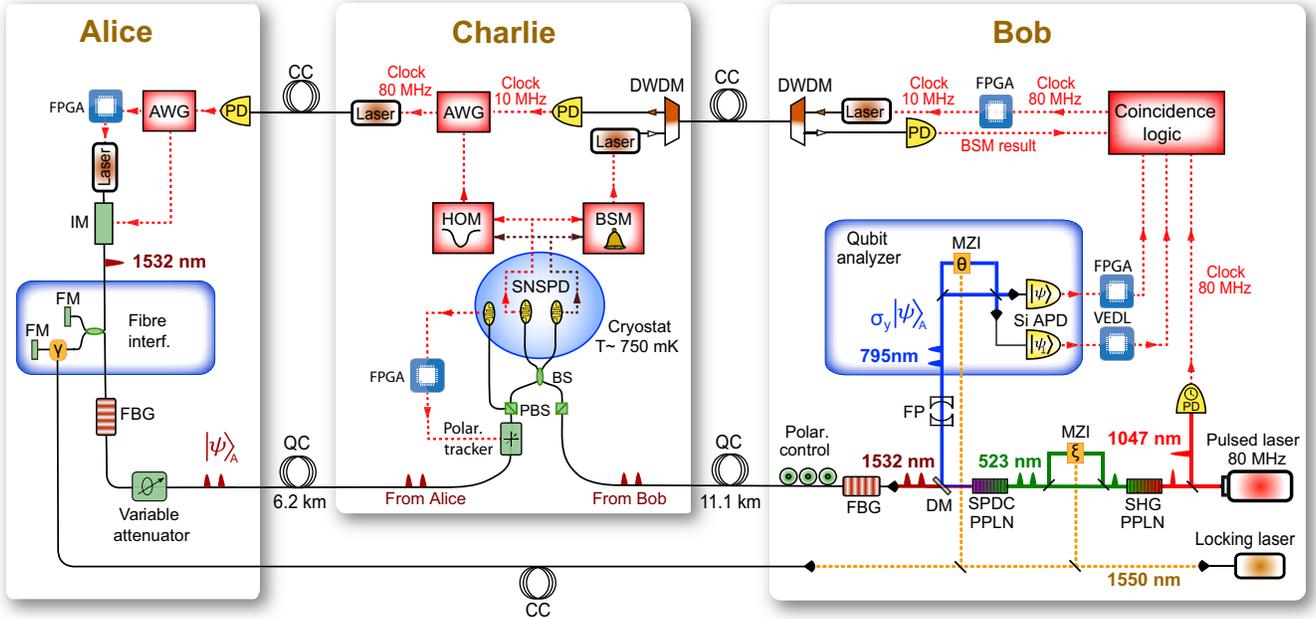}
\caption{\textbf{Schematics of the experimental setup.} 
\textbf{a,} Alice's setup: An intensity modulator (IM) tailors 20 ps-long pulses of light at an 80~MHz rate out of 10~ns-long, phase randomized laser pulses at 1532~nm wavelength. Subsequently, a widely unbalanced fibre interferometer with Faraday mirrors (FM), active phase control (see the Methods sections) and path-length difference equivalent to 1.4 ns~travel time difference creates pulses in two temporal modes or bins. Following their spectral narrowing by means of a 6~GHz wide fibre Bragg grating (FBG) and attenuation to the single-photon level the time-bin qubits are sent to Charlie via a deployed fibre --- referred to as a quantum channel (QC) --- featuring 6~dB loss. 
\textbf{b}, Bob's setup: Laser pulses at 1047~nm wavelength and 6~ps duration from a mode-locked laser are frequency doubled (SHG) in a periodically~poled lithium-niobate (PPLN) crystal and passed through an actively phase-controlled Mach-Zehnder interferometer (MZI) that introduces the same 1.4~ns delay as between Alice's time-bin qubits. Spontaneous parametric down-conversion (SPDC) in another PPLN crystal and pump rejection using an interference filter (not shown) results in the creation of time-bin entangled photon-pairs \cite{Brendel} at 795~and 1532~nm wavelength with mean probability $\mu_\mathrm{SPDC}$ up to 0.06. The 795~nm and 1532~nm (telecommunication-wavelength) photons are separated using a dichroic mirror (DM), and subsequently filtered to 6~GHz by a Fabry-Perot (FP) cavity and an FBG, respectively. The telecom photons are sent through deployed fibre featuring 5.7~dB loss to Charlie, and the state of the 795~nm wavelength photons is analyzed using another interferometer --- again introducing a phase-controlled travel-time difference of 1.4~ns --- and two single photon detectors based on Silicon avalanche photodiodes (Si-APD) with 65\% detection efficiency. 
\textbf{c,} Charlie's setup: A beamsplitter (BS) and two WSi superconducting nanowire single photon detectors \cite{SNSPD} (SNSPD), cooled to 750~mK in a closed-cycle cryostat and with 70\% system detection efficiency, allow the projection of bi-photon states --- one from Alice and one from Bob --- onto the $\ket{\psi^-}$ Bell state. To ensure indistinguishability of the two photons at the BSM, we actively stabilize the photon arrival times and polarization, the latter involving a polarization tracker and polarizing beamsplitters (PBS), as explained in the Methods. Various synchronization tasks are performed through deployed fibres, referred to as classical channels CC, and aided by dense-wavelength division multiplexers (DWDM), photo-diodes (PD), arbitrary waveform generators (AWG), and field-programmable gate-arrays (FPGA), with details in the Methods.
}
\label{fig:setup}
\end{figure*}

An areal map of Calgary, identifying the locations of Alice, Bob and Charlie, is shown in Fig.~\ref{fig:map}, and Fig.~\ref{fig:setup} depicts the schematics of our experimental setup. Alice, located in Manchester (a Calgary neighbourhood), prepares phase-randomized attenuated laser pulses at 1532 nm wavelength with different mean photon numbers $\mu_\mathrm{A}\ll 1$  in various time-bin qubit states \mbox{$\ket{\psi}_A=\alpha\ket{e}+\beta e^{i\phi}\ket{\ell}$}, where $\ket{e}$ and $\ket{\ell}$ denote early and late temporal modes, respectively, $\phi$ is a phase-factor, and $\alpha$ and $\beta$ are real numbers that satisfy $\alpha^2+\beta^2=1$. Using 6.2 km of deployed fibre, she sends her qubits to Charlie, who is located 2.0 km away in a building next to Calgary City Hall.
Bob, located at the University of Calgary (UofC) 6.2 km from Charlie, creates pairs of photons --- one at 1532 nm and one at 795 nm wavelength --- in the maximally time-bin entangled state \mbox{$\ket{\phi^+}=2^{-1/2}(\ket{e,e}+\ket{\ell,\ell})$}. He sends the telecommunication-wavelength photons through 11.1 km of deployed fibre to Charlie, where they are projected jointly with the photons from Alice onto the maximally entangled state \mbox{$\ket{\psi^-}=2^{-1/2}(\ket{e,\ell}-\ket{\ell,e})$}. This leads to the 795 nm wavelength photon at Bob's acquiring the state $\ket{\psi}_B=\sigma_{y}\ket{\psi}_A$, where $\sigma_{y}$ is the Pauli operator describing a bit-flip combined with a phase-flip. In other words, Charlie's measurement results in the teleportation of Alice's photon's state, modulo a unitary transformation, over 6.2 km distance onto Bob's 795 nm wavelength photon.

To confirm successful quantum teleportation, Bob then performs a variety of projective measurements on this photon, whose outcomes, conditioned on a successful BSM at Charlie's, are analyzed using different approaches (see the Methods section for more information on how data is taken). We point out that the 795 nm photons are measured prior to the BSM, thus realizing a scenario where teleportation is achieved \textit{a posteriori} \cite{ma2012a,megedish2013a}. 

\begin{figure}
\centering
\includegraphics[width=\columnwidth]{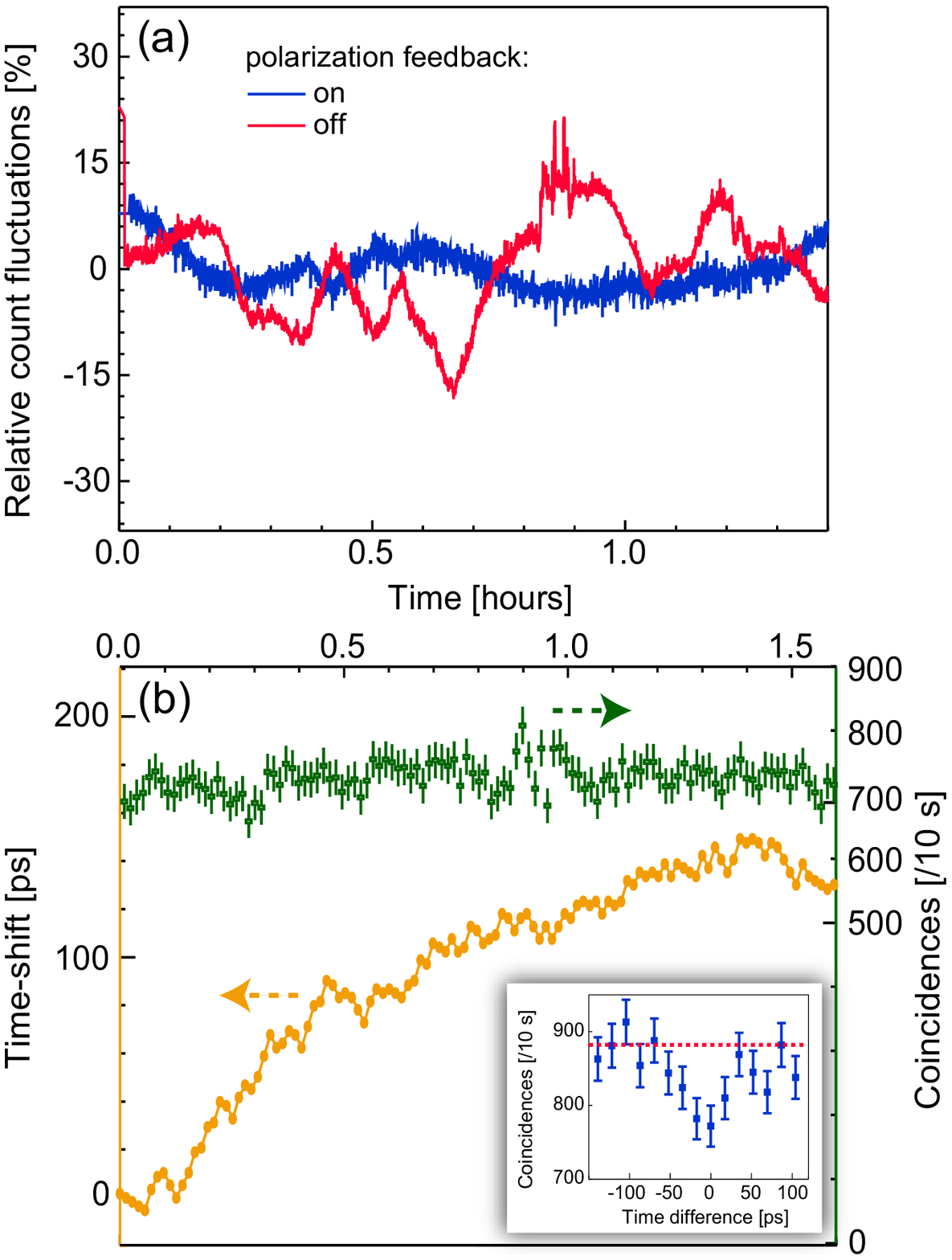}
\caption{\textbf{Indistinguishability of photons at Charlie's.} \textbf{a}, Fluctuations of the count rate of a single SNSPD at the output of Charlie's BS with and without polarization feedback \textbf{b}, Inset: rate of coincidences between counts from SNSPDs as a function or arrival time difference, displaying a Hong-Ou-Mandel (HOM) dip \cite{HOM} when photon-arrival times at the BS are equal. Orange filled circles: The change in the generation time of Alice's qubits that is applied to ensure they arrive at Charlie's BSM at the same time as Bob's. Green empty squares: Coincidence counts per 10 s with timing feedback engaged, showing locking to the minimum of the HOM dip (see Methods). All error bars (one standard deviation) are calculated assuming Poissonian detection statistics.}
\label{fig:indistinguishability}
\end{figure}

The main difficulty in long-distance quantum teleportation is to ensure the required indistinguishability between the two photons subjected to the BSM at Charlie's despite them being created by independent sources and having travelled over several kilometres of deployed fibre. As we show in Fig.~\ref{fig:indistinguishability}, varying environmental conditions during the measurements significantly impact the polarization and arrival times of the photons --- of particular concern being variations of path-lengths differences. Without the active feedback (see the Methods) that our setup performs in an automized way, quantum teleportation would be impossible.  

To verify successful teleportation, first, Alice creates photons in an equal superposition of $\ket{e}$ and $\ket{\ell}$ with a fixed phase, and Bob makes projection measurements onto states described by such superpositions with various phases. Conditioned on a successful BSM at Charlie's, we find sinusoidally varying triple-coincidence count rates with a visibility of ($38 \pm 4$)\% and an average of 17.0 counts per minute. This result alone already represents a strong indication of quantum teleportation: assuming that the teleported state is a statistical mixture of a pure state and white noise, the visibility consistent with the best classical strategy and assuming Alice creates single photons is 33\% \cite{Massar95}. However, here we use this result merely to establish absolute phase references for Alice's and Bob's interferometers. 

\begin{figure}
\centering
\includegraphics[width=\columnwidth]{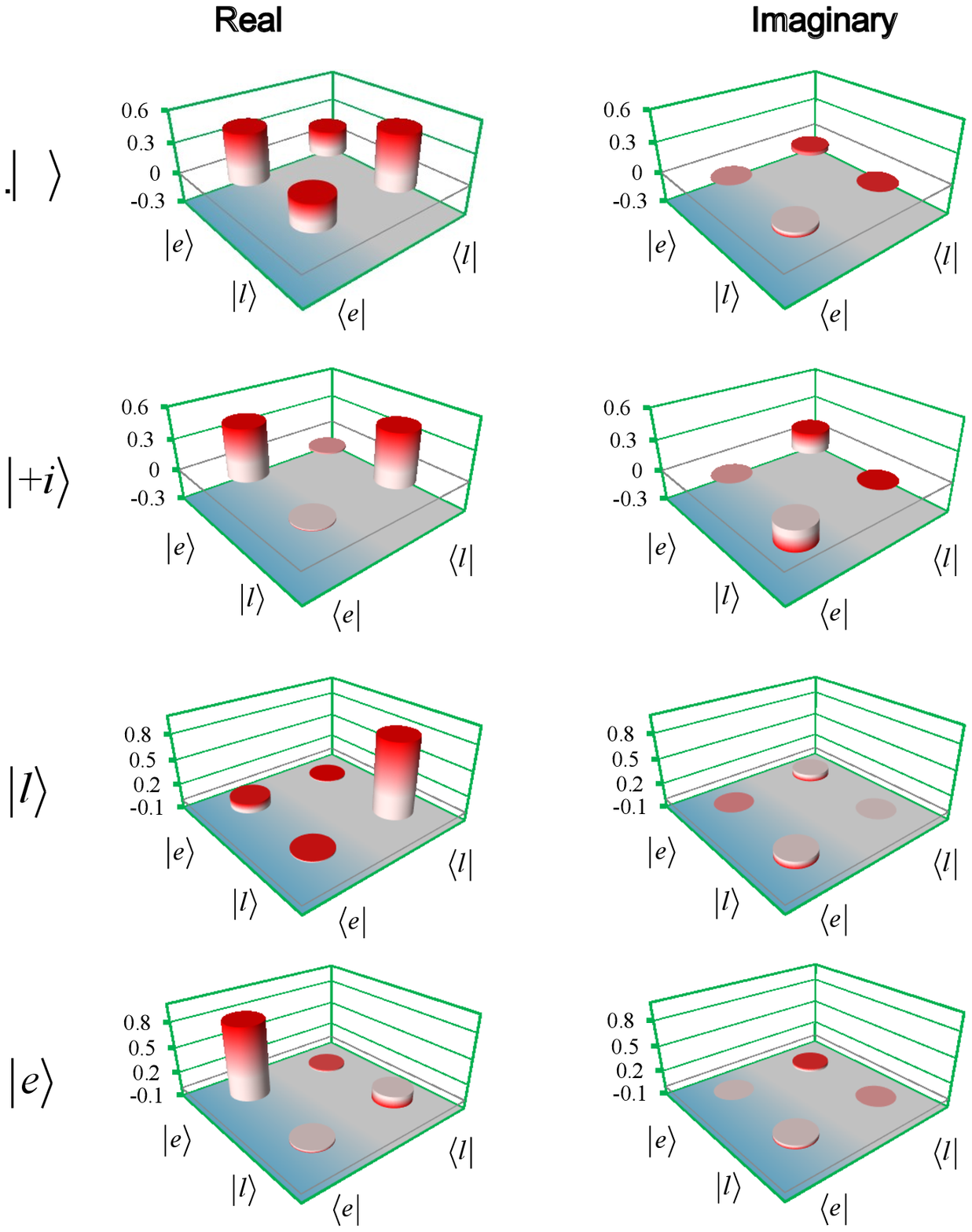}
\caption{\textbf{Density matrices of four states after teleportation.} Shown are the real and imaginary parts of the reconstructed density matrices for four different input states created at Alice's. The mean photon number per qubit is $\mu_\mathrm{A}=0.014$, and the mean photon pair number is $\mu_\mathrm{SPDC}=0.045$. The state labels denote the states expected after teleportation.}
\label{fig:QST}
\end{figure}

\begin{figure}
\centering
\includegraphics[width=\columnwidth]{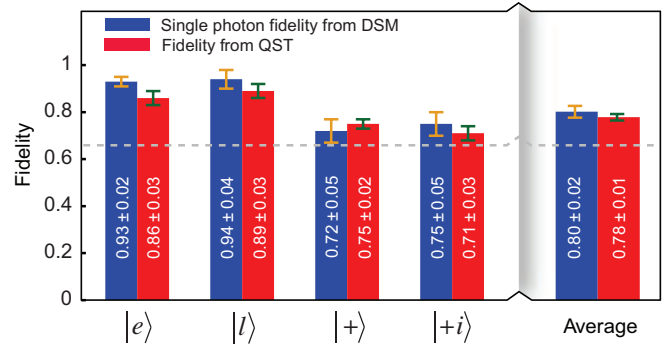}
\caption{Individual and average fidelities of four teleported states with expected (ideal) states, measured using quantum state tomography (QST) and the decoy-state method (DSM). For the DSM we set $\mu_\mathrm{SPDC}=0.06$. Error bars (one standard deviation) are calculated assuming Poissonian detection statistics and using Monte-Carlo simulation.}
\label{fig:fidelity}
\end{figure}

Being able to control the absolute phase values, we can now create photons in, and project them onto, well defined states, e.g. $\ket{e}$, $\ket{\ell}$, $\ket{\pm}\equiv 2^{-1/2}(\ket{e}\pm\ket{\ell})$, and $\ket{\pm i}\equiv 2^{-1/2}(\ket{e}\pm i\ket{\ell})$. This allows us to reconstruct the density matrices $\rho_\mathrm{out}$ of various quantum states after teleportation, and, in turn, calculate the fidelities $F={}_\mathrm{B}\bra{\psi}\rho_\mathrm{out}\ket{\psi}_\mathrm{B}$ with the expected states $\ket{\psi}_\mathrm{B}$. The results, depicted in Figs.~\ref{fig:QST} and ~\ref{fig:fidelity}, show that the fidelity for all four prepared states exceeds the maximum classical value of 66\% \cite{Massar95}. In particular, the average fidelity $\langle F\rangle= \left[F_e+F_l+2(F_++F_{+i})\right]/6$ = (78$\pm$1)\% violates this threshold by 12 standard deviations.

One may conclude that this result shows the quantum nature of the disembodied state transfer between Charlie and Bob. However, strictly speaking, the 66\% bound only applies to Alice's state being encoded into a single photon, while our demonstration, as others before, relied on attenuated laser pulses. To extract the appropriate experimental value, we therefore take advantage of the so-called decoy-state method, which was developed for quantum key distribution (QKD) to assess an upper bound on the error rate introduced by an eavesdropper on single photons emitted by Alice \cite{decoy05, Wang}. Here, we rather use it to characterize how a quantum channel  --- in our case the concatenation of the direct transmission from Alice to Charlie and the teleportation from Charlie to Bob --- impacts on the fidelity of quantum states encoded into individual photons \cite{Sinclair15}. Towards this end, we vary the mean number of photons per qubit emitted at Alice's between three optimized values, $\mu_\mathrm{A}\in \{0, 0.014, 0.028\}$, and calculate error rates and transmission probabilities for each value independently. The results, also depicted in Fig.~\ref{fig:fidelity}, show again that the fidelities for all tested states exceed the maximum value of 2/3 achievable in classical teleportation. We note the good agreement between the measured values and those predicted by our model developed (inspired by \cite{felix14}) that takes into account various, independently characterized system imperfections (no fit). This allows us to identify that deviations of the measured fidelities from unity  --- i.e. from ideal teleportation --- are mostly due to remaining distinguishability of the two photons subjected to the BSM at Charlie's, followed by multi-pair emissions by the pair-source. Finally, by averaging the single-photon fidelities over all input states, weighted as above, we find  $\langle F^{(1)}\rangle \geq (80\pm 2)$\% --- as before significantly violating the threshold between classical and quantum teleportation.

Our measurements establish the possibility for quantum teleportation over many kilometres in the important mid-point configuration --- as is required for extending the distance of quantum communications using quantum repeaters. We emphasize that both photons travelling to Charlie are at telecommunication wavelength, making it possible to extend the Alice-Bob distance from its current value of 8 kilometres by at least one order of magnitude. This corresponds to the distance of an elementary link, which includes teleportation of entanglement, at which communication links based on spectrally multiplexed quantum repeaters start to outperform direct qubit transmission \cite{Sinclair15,krovi15}. 
We also note that the 795 nm photon, both in terms of central wavelength as well as spectral width, is compatible with quantum memory for light --- a key element of a quantum repeater --- in cryogenically-cooled thulium-doped crystals, in particular Tm:YGG, whose spectral acceptance exceeds the bandwidth needed for most practical applications \cite{Thiel14}, and using which we have recently demonstrated storage times of up to 100 $\mu$sec \cite{Sinclair16}. Finally, we note that quantum teleportation involves the interesting aspect of Alice transferring her quantum state in a disembodied fashion to Bob without him ever receiving any physical particle. In other words, Bob is only sending photons (all of them members of an entangled pair) and thus better able to protect his system from any outside interference, e.g. from an adversary. This points to similar considerations of security as measurement-device-independent QKD \cite{Lo}, albeit in a more flexible quantum network setting that could allow, e.g., distributed quantum computing \cite{kimble}. These key features make our demonstration an important step towards long-distance quantum communication, and ultimately a global quantum Internet.

We note that, a related experimental demonstration has been reported in a concurrent manuscript \cite{Pan16}.

\hspace{1cm} 

\noindent
\textbf{Acknowledgements}\\
The authors thank Tyler Andruschak and Harpreet Dhillon from the City of Calgary for providing access to the fibre network and for help during the experiment, Vladimir Kiselyov for technical support, and Pascal Lefebvre for help with aligning the entangled photon pair source. This work was funded through Alberta Innovates Technology Futures (AITF), the National Science and Engineering Research Council of Canada (NSERC), and the Defense Advanced Research Projects Agency (DARPA) Quiness program (contract no. W31P4Q-13-l-0004). WT furthermore acknowledges funding as a Senior Fellow of the Canadian Institute for Advanced Research (CIFAR), and VBV and SWN acknowledge partial funding for detector development from the Defense Advanced Research Projects Agency (DARPA) Information in a Photon (InPho) program. Part of the detector research was carried out at the Jet Propulsion Laboratory, California Institute of Technology, under a contract with the National Aeronautics and Space Administration.
\\

\noindent
\textbf{These authors contributed equally to this work}\\
Raju Valivarthi, Marcel.li {Grimau Puigibert}, Qiang Zhou, and Gabriel H. Aguilar.\\

\noindent
\textbf{Author contributions}\\
The SNSPDs were fabricated and tested by VBV, FM, MDS, and SWN. The experiment was conceived and guided by WT. The setup was developed, measurements were performed and the data analyzed by RV, MGP, QZ, GHA, and DO. The manuscript was written by WT, RV, MGP, QZ, GHA, and DO.\\


\noindent
\textbf{Competing financial interests}\\
The authors declare no competing financial interests.

\newpage
\section*{Methods}

\subsection{Synchronization}
For the following discussion, please refer to experimental setup outlined in Fig.~\ref{fig:setup}. Charlie is connected via pairs of optical fibres both to Alice and to Bob. In each pair, one fibre --- referred to as the quantum channel (QC) --- carries single photons, while the other --- referred to as the classical channel (CC) --- distributes various strong optical signals whose purpose will be described in the following. In addition, Alice and Bob are directly connected via an optical fibre that transmits narrow-line-width laser light at 1550~nm in order to lock all interferometers. This is crucial for maintaining a common phase reference for the qubit states generated at Alice's and Bob's, and analyzed at Bob's. In each interferometer, the power of the locking laser in one output arm (measured on a classical detector) is used to derive a feedback signal  to a piezo-element that regulates the path-length difference of the interferometer to maintain a fixed phase. Additionally, all interferometers are kept in temperature controlled boxes. 

The master clock for all devices is derived from detection of the mode-locked laser pulses (80~MHz) and converted back into an optical signal for distribution through the CC via Charlie to Alice.

\subsection{Stabilization to ensure photon indistinguishability} 
For a successful BSM, the photons arriving at Charlie's from Alice and Bob need to be indistinguishable despite being generated by independent and different sources, and having travelled through several kilometres of deployed fibre. To ensure that the photons have the same spectral profile, they are sent through separate, temperature-stabilized fibre Bragg gratings (FBG) that narrow their spectra to 6~GHz. The spectral overlap of the FBGs at Alice and Bob needs to be set only once. However, due to temperature-dependent properties of fibre such as birefringence and length, the polarization and arrival time of the photons change with external environmental conditions, making it difficult to implement the BSM in a real-world environment. Towards this end, we apply feedback mechanisms to compensate for drifts in polarization and arrival time.

\subsubsection{Timing}
The short duration of our photons ($\sim70$~ps) prevents us from using the SNSPDs (featuring a time jitter of $\sim150$ ps) to directly determine their arrival times with the required precision to adjust the difference to zero. Instead, we compensate for  arrival-time drifts with the novel approach of observing the degree of quantum-interference (Hong-Ou-Mandel or HOM effect \cite{HOM}) of the photons. The signals from the two SNSPDs (which are used to perform the BSM) are also sent to a HOM analyzing unit (see Fig.~\ref{fig:setup}) that monitors the number of coincidences between detections corresponding to either both photons arriving in an early time bin mode, or both in a late bin. As shown in the inset of Fig.~\ref{fig:indistinguishability}b, the HOM interference causes photon bunching and thus the coincidences to be reduced when the photons arrive at the beam splitter at the same time.  Hence, to counteract the drift in travel time of the photons, we vary the qubit generation time at Alice with a precision of about $\sim$4~ps to keep the coincidence count rate at the minimum value of around 750 per 10 sec., as shown in Fig.~\ref{fig:indistinguishability}b. In practice Alice's time-shift is triggered at Charlie's by shifting the phase of the master clock signal that he forwards to Alice. Fig.~\ref{fig:indistinguishability}b, shows that, during a typical 1.5~hour measurement, we apply a time shift of $\sim$200 ps to compensate drifts in timing. Since the shift is larger than the duration of the photons, the teleportation protocol would fail without the active stabilization.

\subsubsection{Polarization}
Because photons from Alice and Bob pass through polarizing beam-splitters (PBS) at Charlie's, their polarization indistinguishability is naturally satisfied. However,  correct photon polarizations must be set and maintained to maximize the transmission through the PBSs, or else the channel loss will vary over time. In our system, the QC between Bob and Charlie experiences only a slow drift, which allows for manual compensation using a polarization controller --- located at Bob's --- once a day. However, an automated polarization feedback system is required for the channel between Alice and Charlie, which drifts significantly on the time-scale of the experiment. To that end, we monitor the count rate of an additional SNSPD, located in the reflection port of the PBS at Charlie's, with a field-programmable gate-arrays (FPGA) so as to generate a feedback signal that minimizes the rate by adjusting the polarization by means of a polarization tracker (also located at Charlie's). As seen in Fig.~\ref{fig:indistinguishability}, the intensity fluctuations in 1.5 hours (a typical time scale to acquire results for one qubit setting) are limited to 5$\%$ with feedback, and to about 15$\%$ without feedback.

\subsection{Data collection} 
Using Alice's qubits and the 1532 nm-members of the entangled pairs, Charlie performs $\ket{\psi^{-}}$ Bell-state projections. Such a projection occurs when one SNSPD detects a photon arriving in the early time-bin and the other SNSPD records a photon in the late time-bin. Successful Bell-state projection measurements are communicated via the CC and using classical laser pulses to Bob's, who converts them back to electrical signals. Each signal is then used to form a triple coincidence with the detection signal of the 795~nm wavelength photon exiting Bob's qubit analyzer. Towards this end, the latter is delayed using a variable electronic delay-line (VEDL) implemented on an FPGA by the time it takes the 1532~nm entangled photon to travel from Bob to Charlie plus the travel time of the BSM signal back to Bob.

\end{document}